\documentclass[12pt]{article}
\pdfoutput = 1
\usepackage{graphicx}
\usepackage{adjustbox}
\usepackage{fancybox}
\usepackage{dcolumn}
\usepackage{multirow}
\usepackage{multicol}
\usepackage{amssymb}
\usepackage{amsthm}
\usepackage{amsfonts,amsmath}
\usepackage{appendix}
\usepackage{tikz} 
\usetikzlibrary{fit,positioning}
\usepackage[section]{placeins}
\usepackage{cite}
\usepackage{lineno}
\usepackage{caption}
\usepackage{subcaption}
\usepackage{float}
\usepackage{graphicx}
\usepackage{hyperref}
\usepackage{colortbl}
\usepackage{authblk}
\usepackage{array}
\usepackage{adjustbox}
\usepackage{cite}
\usepackage{multirow}
\usepackage{nameref,hyperref}
\usepackage{multicol}

\usepackage{color, colortbl}
\usepackage{authblk}
\definecolor{Gray}{gray}{0.9}
\definecolor{Gray1}{gray}{0.7}
\usepackage{adjustbox}
\usepackage{tabularx,colortbl}

\title{Projecting the Impact of Covid-19 Variants and Vaccination Strategies in Disease Transmission using a Multilayer Network Model in Costa Rica}

\author[1,3, +]{Yury E. Garc\'ia}
\author[4, +]{Gustavo Mery}
\author[1, +]{Paola V\'asquez}
\author[2, +]{Juan G. Calvo}
\author[2, +]{Luis A. Barboza}
\author[5, +]{Tania Rivas}
\author[2, +]{Fabio Sanchez}

\affil[1]{Universidad de Costa Rica, Centro de Investigaci\'on en Matem\'atica Pura y Aplicada, San Jos\'e, Costa Rica}

\affil[2]{Universidad de Costa Rica, Centro de Investigaci\'on en Matem\'atica Pura y Aplicada - Escuela de Matem\'atica, San Jos\'e, Costa Rica}

\affil[3]{Department of Public Health Sciences, 
University of California Davis, CA, USA}

\affil[4]{Pan America Health Organization, World Health Organization, San José 10102, Costa Rica.}

\affil[5]{Ministry of Health, San José 10102, Costa Rica}

\affil[*]{yury.garciapuerta@ucr.ac.cr}

\affil[+]{these authors contributed equally to this work}

\date{}
\begin{document}
\maketitle

\begin{abstract}
For countries starting to receive steady supplies of vaccines against SARS-CoV-2, the course of Covid-19 for the following months will be determined by the emergence of new variants and successful roll-out of vaccination campaigns. To anticipate this scenario, we used a multilayer network model developed to forecast the transmission dynamics of Covid-19 in Costa Rica, and to estimate the impact of the introduction of the Delta variant in the country, under two plausible vaccination scenarios, one sustaining Costa Rica's July 2021 vaccination pace of 30,000 doses per day and with high acceptance from the population and another with declining vaccination pace to 13,000 doses per day and with lower acceptance. Results suggest that the introduction and gradual dominance of the Delta variant would increase Covid-19 hospitalizations and ICU admissions between $35\%$ and $33.25\%$, from August 2021 to December 2021, depending on vaccine administration and acceptance. In the presence of the Delta variant, new Covid-19 hospitalizations and ICU admission would experience an average increase of $24.26\%$ and $27.19\%$ respectively in the same period if the vaccination pace drops. Our results can help decision-makers better prepare for the COVID-19 pandemic in the months to come. 
\end{abstract}

\section*{Introduction}

In the fight against the Covid-19 pandemic, the remarkable time frame at which scientists achieved to develop safe and effective vaccines~\cite{kim2020covid,li2021comprehensive} has provided a crucial tool in the global effort to reduce the widespread health, economic, and social disruption caused by the emergence of the severe acute respiratory syndrome coronavirus 2 (SARS-CoV-2). As of July 2021, just 18 months after the genetic sequencing of SARS-CoV-2~\cite{wang2020genetic}, a total of 21 vaccines have been granted emergency use authorization by countries around the world~\cite{Vaccines57:online}, while 108 vaccine candidates are at different phases of clinical development~\cite{COVID19v90:online}. 

Despite these extraordinary achievements, worldwide vaccination programs are facing challenges, such as supply shortages~\cite{wouters2021challenges}, mainly in low-to-middle-income countries (LMIC)~\cite{acharya2021access}, and low acceptance rates, mostly in their high-income counterparts~\cite{arce2021covid,sallam2021covid}.

Furthermore, the world is facing the ominous threats of rapidly appearing SARS-CoV-2 variants~\cite{abdool2021new,walensky2021sars,Tracking0:online}. Several variants of concern have already increased the transmissibility of the virus~\cite{davies2021estimated} and the severity of cases among younger population groups~\cite{nyberg2021risk,challen2021risk}. Moreover, these variants have shown evidence of reduced effectiveness of current vaccines, at least in lowering disease transmission~\cite{bernal2021delta,COVID19v12:online}. We have already started to hear calls for third doses to fight more aggressive variants~\cite{callaway2021covid}, which are more likely to appear if vaccine supplies remain scarce in large parts of the world.

In light of this, unless worldwide vaccine distribution becomes more homogeneous, the current vaccine shortages in large parts of the world will perpetuate SARS-CoV-2 circulation, which ensures the relentless appearance of new variants~\cite{COVIDvar91:online}. Something similar is expected in countries where, despite having enough vaccines, large pockets of the population remain unvaccinated due in part to misinformation and ineffective health promotion and communication strategies~\cite{loomba2021measuring,van2021inoculating}.

Under this context, the development of new models is providing health authorities and decision-makers around the world an evidence-based tool to inform, guide, update and adapt public health interventions to their unique circumstances and available resources. These models can be used as a tool to forecast potential transmission scenarios and the associated healthcare demand under different vaccination scenarios~\cite{moore2021modelling,matrajt2021optimizing,paltiel2021speed,coudeville2021potential,romero2021public,tetteh2020covid,giordano2021modeling,usherwood2021model} and circulating variants~\cite{rella2021rates,caldwella2021vaccines}.

In line with these efforts, the current study was developed by implementing a tailored multilayer network model for Covid-19 in Costa Rica. The objective of the analysis was to project and assess the impact of the introduction of the more transmissible B.1.617.2 (Delta) variant in the country. We compared the effects of two different vaccine administration scenarios on the number of SARS-CoV-2 infections, deaths, and Covid-19 related hospital admissions from August 2021 to December 2021. We were also able to estimate the direct base cost of providing hospital care. Since vaccination pace depends on the availability of the vaccine and the acceptance by the population, vaccination scenarios in the model considered both the speed of vaccine administration and percentages of vaccine hesitancy from the population.

This study considered the information available by July 31, 2021. As of that date, Costa Rica had only administered the Pfizer-BioNTech (BNT162B2) and AstraZeneca (ChAdOx1-s) vaccines. The vaccination program in the country began on December 24, 2020. For its distribution, health authorities notified a vaccination strategy which divided the population into five groups taking into consideration the age of the population, work-related risk of exposure and preexisting medical conditions. These groups and their opportunity to receive the jabs have been adjusted and revised frequently by the National Vaccination Committee as new evidence and the supply of vaccines became accessible~\cite{Lineamie32:online}. By May 18, 2021, the majority of the vaccinated population had received the 21-days interval scheme with BNT162B2, as the first ChAdOx1-s vaccines entered the country during the second week of April. After evidence of increasing the intervals between doses arose~\cite{parry2021extended} the government announced a 12-week interval between doses for both the BNT162B2 and ChAd0x1-s vaccines for people under 57 years of age. Thus, we considered that all vaccinated individuals received their second dose 12 weeks after the first dose as of May 18, 2021.

By July 20 2021, the country had increased its vaccination capacity and it was inoculating at a rate of 30,000 doses per day, reaching $38.4\%$ of the total population covered with one dose and $16.1\%$ fully vaccinated by that same date. However, several counties in the national territory were showing lower vaccination acceptance, and vaccine hesitation appear higher among younger and healthier individuals, which threatens to mirror declines in vaccination rates observed in several other countries. 

On the other hand, by July the country was facing the imminent arrival of the most transmissible Delta variant. Earlier in 2021, the country saw its worse Covid-19 crisis between May and early June, period that coincides with the introduction of the Alpha and even more dominant Gamma variants. Cases by late June and early July were rapidly dropping at last, as Covid-19-related hospitalizations and deaths were too. But countries around the world were facing rapid resurgences in cases and hospital admission due to the newest Delta variant \cite{WHOCoron71:online}. It was uncertain how the introduction of this more aggressive pathogen will affect the pandemic in the country by the end of 2021, when vaccination coverage was expected to reach numbers which were once thought to provide herd immunity. 
\section*{Methods}

\subsection*{Data}

Publicly available information of Covid-19 regarding daily reported cases, hospitalization, and ICU admissions was obtained from the official website of the Ministry of Health of Costa Rica~\cite{Lineamie32:online}. We used demographic data of the 5,163,038 inhabitants of the country to establish the contact network. This data included population by age, canton of residence and workplace, as reported by the National Institute of Statistics and Census (INEC)~\cite{Bienveni11:online}. 

For the initial conditions of the model we incorporated the number of confirmed Covid-19 cases from March 6, 2020 (day the first case was reported in the country~\cite{Lineamie32:online}), to June 28, 2021, including the age and canton of residence of the patients as notified by the health authorities. During this period, the cumulative number of detected cases ascended to 403,511 with a case fatality rate of $1.3\%$~\cite{Lineamie32:online}. Parameters related to hospitalization that involve duration of stay, percentage of hospital admissions by age group and hospitalization costs were provided by the Social Security Fund of Costa Rica (CCSS), public entity in charge of both general and specialized medical care for the Costa Rican population. According to data from CCSS, the estimated base cost of hospital bed per day, (which doesn't include diagnosis procedures or treatment) is approximately 707,540 colons (1,137 USD) in the medical ward and 902,641 colons (1,450 USD) in an ICU ~\cite{ccstarifa}. It is worth stressing that hospital charges in reality are substantially higher for Covid-19 patients when all procedures, specialist care and exams are considered, and that cost varies with length of stay.

We also used the weekly number of administered Covid-19 vaccines as reported by CCSS from December 24, 2020 to July 28, 2021~\cite{COVID19C0:online}. According to health authorities, the target population to be vaccinated in Costa Rica is 4,274,344, a $83\%$ of the total population which includes all residents that are 12 years and older. During this period, $55\%$ of the target population had received the first dose, while $19\%$ received the second dose.

As for the percentage of vaccine hesitance, we took into account the available information of administrated vaccines by age group as of July 2021 in Costa Rica, as well as information reported from previous studies, where a median of vaccine acceptance of $78\%$ has been observed in LMICs~\cite{solis2021covid}. 
\subsection*{Model}

We extended a multilayer, temporal and stochastic \textit{network model} first developed to simulate the spread of the SARS-CoV-2 in Costa Rica. A complete description and implementation is presented in~\cite{calvo2021multilayer}. The scenarios were constructed to include different vaccine administration speeds, the introduction of the Delta variant, and population resistance to available Covid-19 vaccines in the country. 
 
The multilayer structure of the model incorporates three layers representing people living in the same house, friends or coworkers, and sporadic contacts. The probability of transmission in each layer is different, and while the contacts of the household layer remain fixed, the other two layers have variations according to public health measures and changes in social behaviour. The model contemplates ten mutually exclusive compartments: susceptible, latent, diagnosed (infectious) and undiagnosed (infectious), hospitalized, individuals admitted to intensive care, recovered, successfully vaccinated with one dose, fully vaccinated, and Covid-19 deceased individuals.

To include the difference in disease severity among age groups~\cite{kang2020age}, vaccination strategies implemented by the Costa Rican health officials, and vaccine hesitance, the model is divided into three age classes: children (0-18 years), adults (19-64 years), and elderly people (+65 years). Assumptions related to disease progression and transmission characteristics remain the same as the original model~\cite{calvo2021multilayer}. However, new hypotheses were introduced to forecast the effects of the different vaccination scenarios.

Individuals to be vaccinated are randomly chosen from the susceptible and recovered compartments. From December 24, 2020, to April 22, 2021, the model incorporated the condition that all recovered individuals would be vaccinated 90 days after their recovery. However, starting from April 23, 2021, this assumption was changed, after health authorities announced that individuals who have already been infected could be vaccinated regardless of the time since their recovery~\cite{Lineamie32:online}.

Given the available evidence, the model takes into account vaccine effectiveness for the Pfizer-BioNTech and AstraZeneca vaccines. Studies have shown that after the first dose with either of these two vaccines (pre-introduction of the Delta variant), protection against symptomatic illness ranges from $55\%$-$70\%$, while vaccine effectiveness against hospitalization has been reported to be $75\%$-$85\%$~\cite{bernal2021effectiveness,COVID19v59:online}. Given these levels of protection, our model does not differentiate between administered vaccines. As for vaccine effectiveness against the Delta variant, we assumed a reduction in protection against symptomatic disease after both first and second dose as reported in~\cite{bernal2021delta,COVID19v59:online}. 

\subsection*{Scenarios}

We developed two scenarios considering the maximum and minimum average of daily first doses during the initial seven months of the vaccination campaign in Costa Rica. The first vaccine scenario contemplates an accelerated vaccination rate with 30,000 daily doses and a high percentage of acceptance by the population, $85\%$ for those aged between 19 and 64 and $95\%$ for people older than 65. The second scenario consists of having a decelerated vaccination rate with 13,000 daily doses and a low acceptance by individuals to receive the vaccine, $70\%$ for those aged between 19 and 64 and $90\%$ for people over 65 years old. 

Vaccine administration speed is implemented in two contexts: when circulating variants up to the first half of 2021 continue to dominate for the remainder of the year, and when the Delta variant is detected and gradually becomes dominant.

For the case where the Delta variant is detected during mid-July and becomes dominant two months after, we assumed a vaccine efficacy against symptomatic infections to be $55\%$ at the end of August and $45\%$ from mid-October for those with a single dose. Its important to note, that these percentages consider that other variants will continue to circulate therefore we did not decrease vaccine protection against symptomatic illness for the Delta variant to the $30\%$-$35\%$ reported in literature~\cite{bernal2021delta,COVID19v59:online}. For those fully vaccinated, these percentages are $82\%$ at the end of August and $72\%$ from mid-October onward. No changes are made in the percentages of hospitalization. Recent studies showed that two doses of Pfizer-BioNTech (BNT162B2) or AstraZeneca (ChAdOx1-s) have similar effectiveness in preventing hospitalizations for the Delta coronavirus variant as they are against the previously dominant variants~\cite{bernal2021delta,COVID19v59:online}.

Vaccines are also administrated according to the age of the population, starting with people over 65 years old, then people between 19 and 64, and finally between 12 and 18. Initially, $85\%$ of the total doses were given to people over 65 years old and $15\%$ to people aged 19 to 64 years. When the target percentage of people in each age group is reached, $100\%$ of the vaccines are administrated in the next available age group. All simulations assume an efficacy against symptomatic infection of $63\%$ after the first dose and $90\%$ after the second one. These percentages change over time in the scenarios considering the introduction of the Delta variant. In the model, there are no differences in the prevention of severe symptoms between people with one or two doses. We assume that the probability of hospitalization among those vaccinated decreases as the total number of fully vaccinated people increases. The probability of hospitalization is $20\%$ until reaching $40\%$ of the vaccinated target population, then decreases by $15\%$ when $60\%$ of the vaccinated target population has a complete scheme. From that moment on, the probability is $10\%$. 

As vaccination rates increase, we assume that individuals begin to feel safe, this leads to a decrease in the percentage of people adhering to the no pharmaceutical interventions. Furthermore, parameters related to these measures within the model are varied when $60\%$ of the target population are fully vaccinated. All scenarios are simulated until December 31, 2021. A summary of the assumptions are presented in Table~\ref{tab:scenarios}.

\begin{table}[!ht]
\centering
\begin{tabular}{|>{\raggedleft\arraybackslash}m{5.3cm}|>{\centering\arraybackslash}m{3.7cm}|>{\centering\arraybackslash}m{3.7cm}|}
\hline
\rowcolor{Gray}
Parameter & Scenario 1 & Scenario 2 \\ \hline
Daily doses from August 2, 2021 & 13,000 & 30,000 \\ \hline
\% individuals over 65 that are not vaccinated  & 10 & 5 \\ \hline
\% individuals from 19 to 64 that are not vaccinated  & 30 & 15 \\ \hline
\% social distancing from July 12, 2021 & \multicolumn{2}{l|}{\begin{tabular}[l]{@{}l@{}}\{55, 60, 65, 70\} initially\\ \{40, 45, 50, 55\} when 60\% of the target \\       population is fully vaccinated\end{tabular}} \\ \hline
\% individuals following NPI$^*$ & \multicolumn{2}{l|}{\begin{tabular}[l]{@{}l@{}}\{50, 55, 60, 65, 70\} when 60\% of the target\\ population is fully vaccinated\end{tabular}} \\ \hline
\rowcolor{Gray}
Parameter & Before Delta & Delta variant \\ \hline
\% protection against symptomatic infection, first dose & 63 &  \multicolumn{1}{l|}{\begin{tabular}[l]{@{}l@{}} 55 after August 30, 2021\\ 45 after October 15, 2021\end{tabular}}\\ \hline
\% protection against symptomatic infection, second dose & 90 &  \multicolumn{1}{l|}{\begin{tabular}[l]{@{}l@{}} 82 after August 30, 2021\\ 72 after October 15, 2021
\end{tabular}}  \\ \hline
\end{tabular}\\
\footnotesize{$^*$Personal protection measures such as mask use, washing hands and keeping social distance.}
\caption{Summary of scenarios.}
\label{tab:scenarios}
\end{table}

\section*{Results}

\begin{figure}[!ht]
\centering
\includegraphics[scale=0.4]{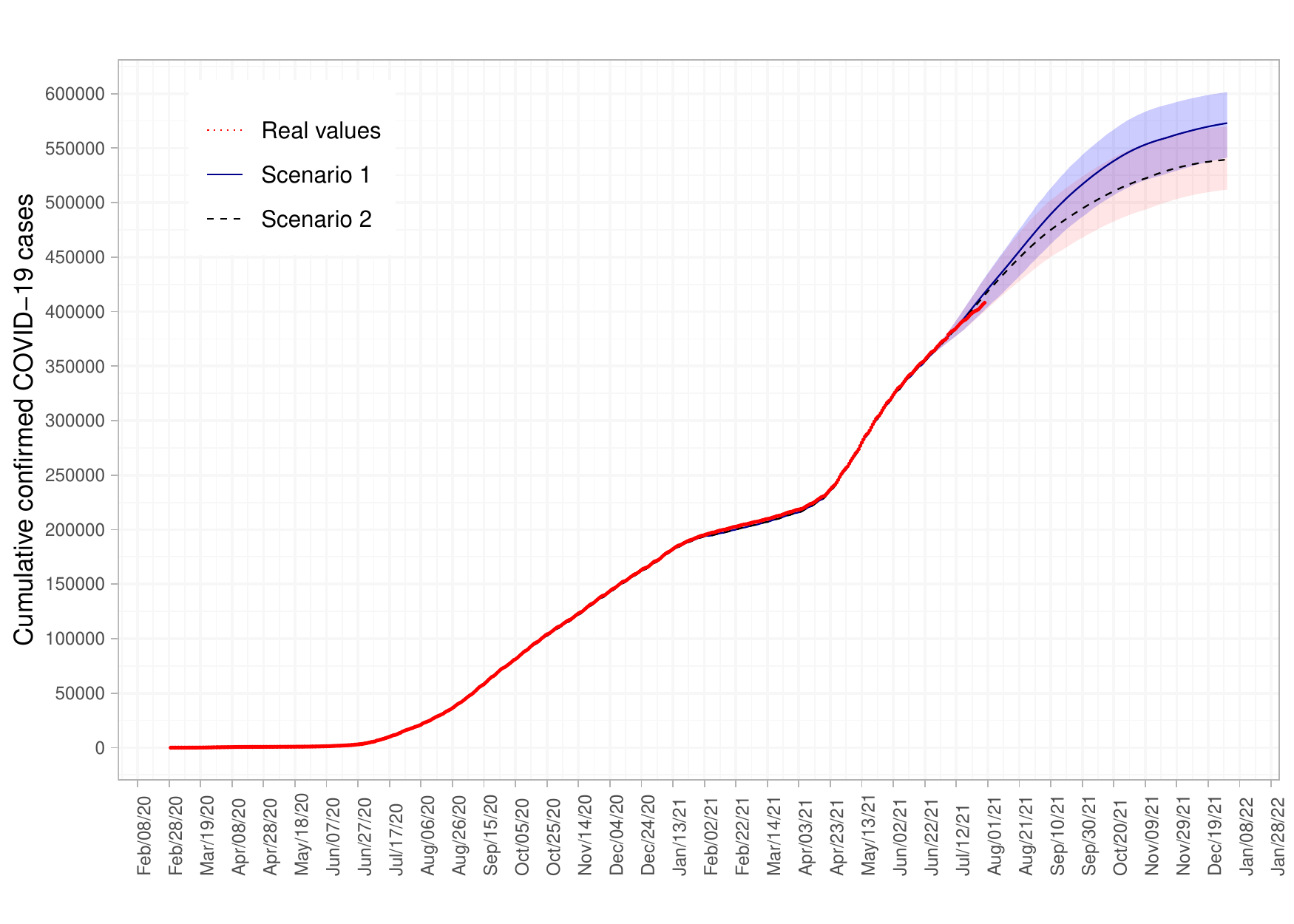}
\includegraphics[scale=0.4]{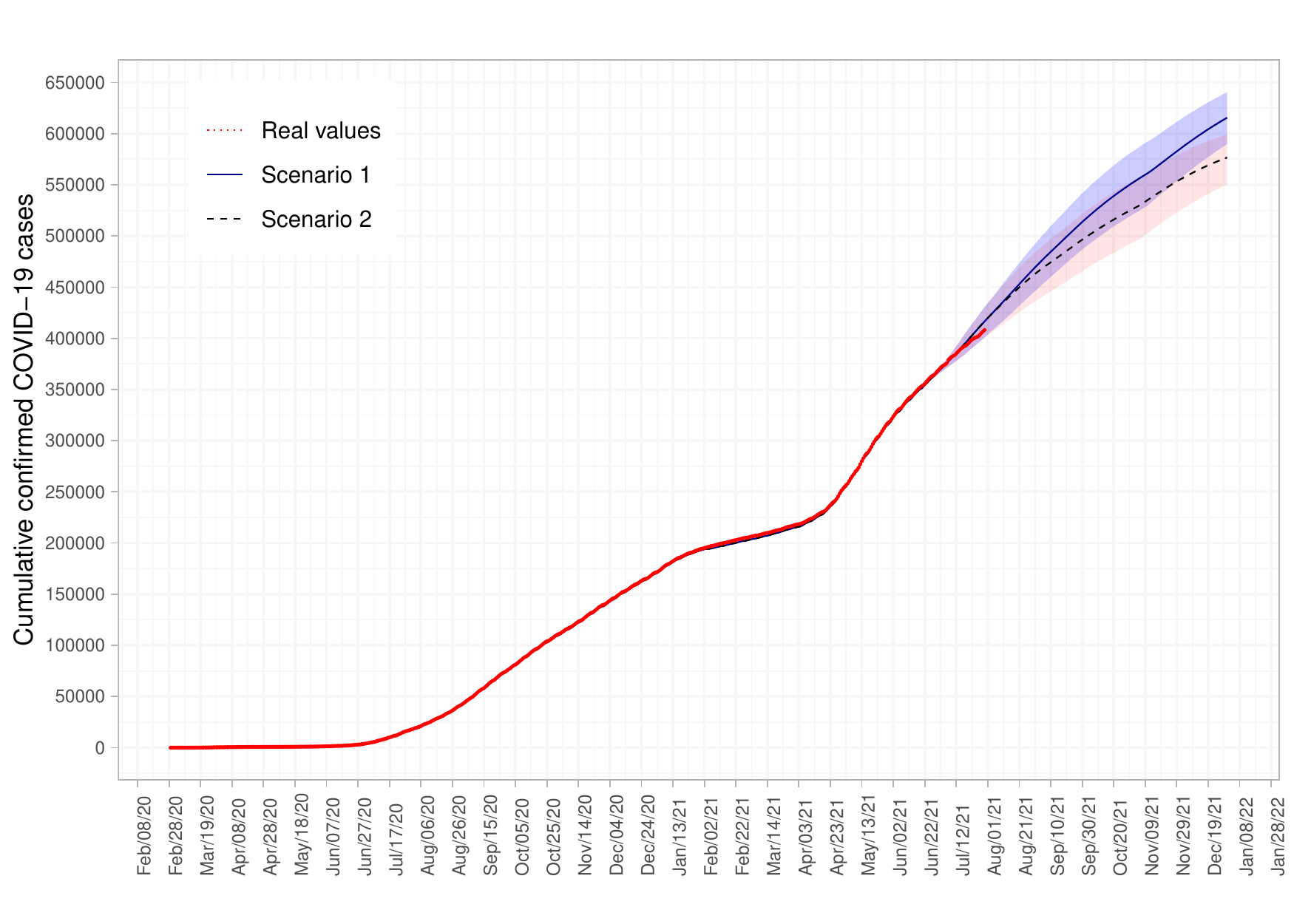}\\
\includegraphics[scale=0.4]{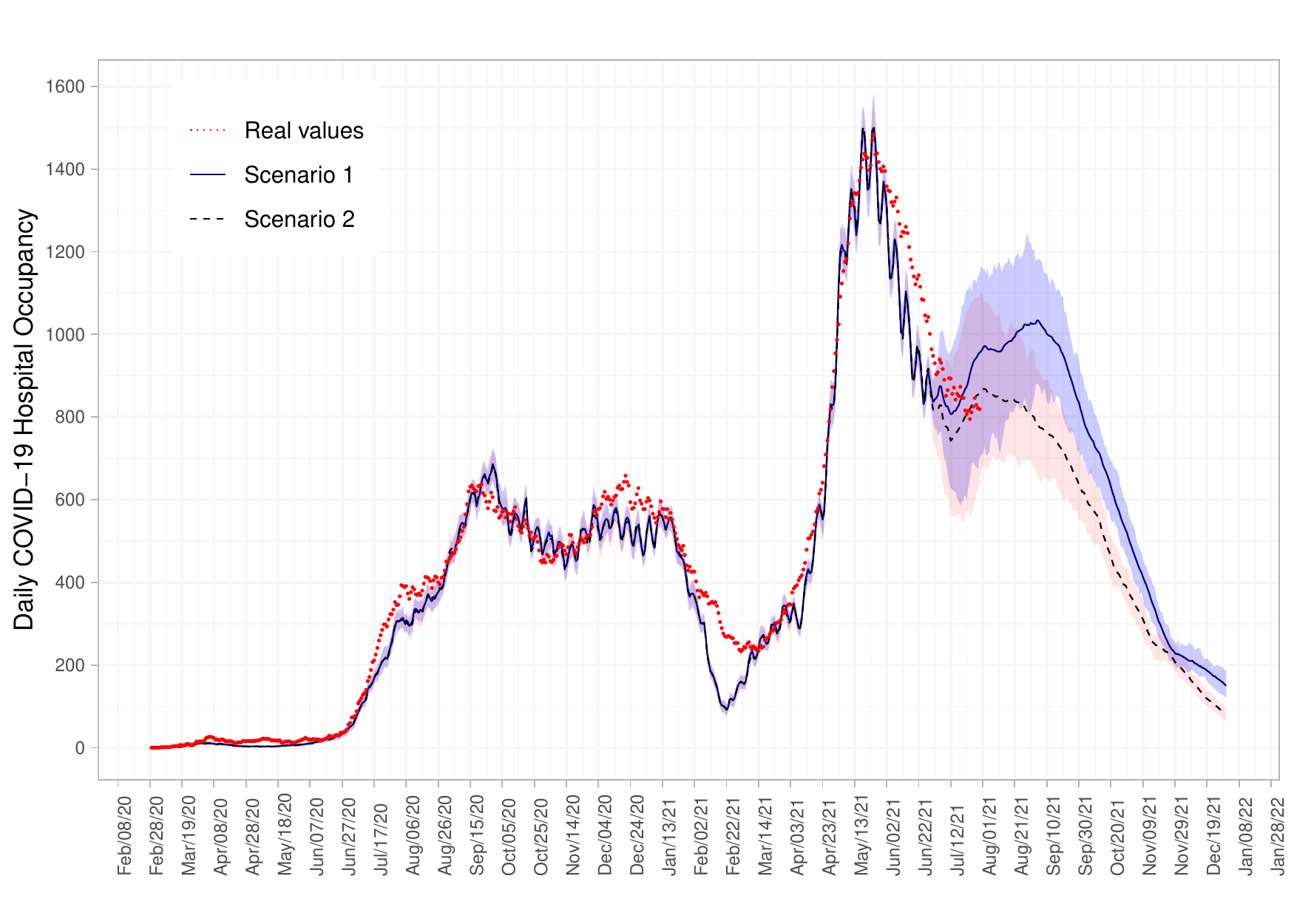}
\includegraphics[scale=0.4]{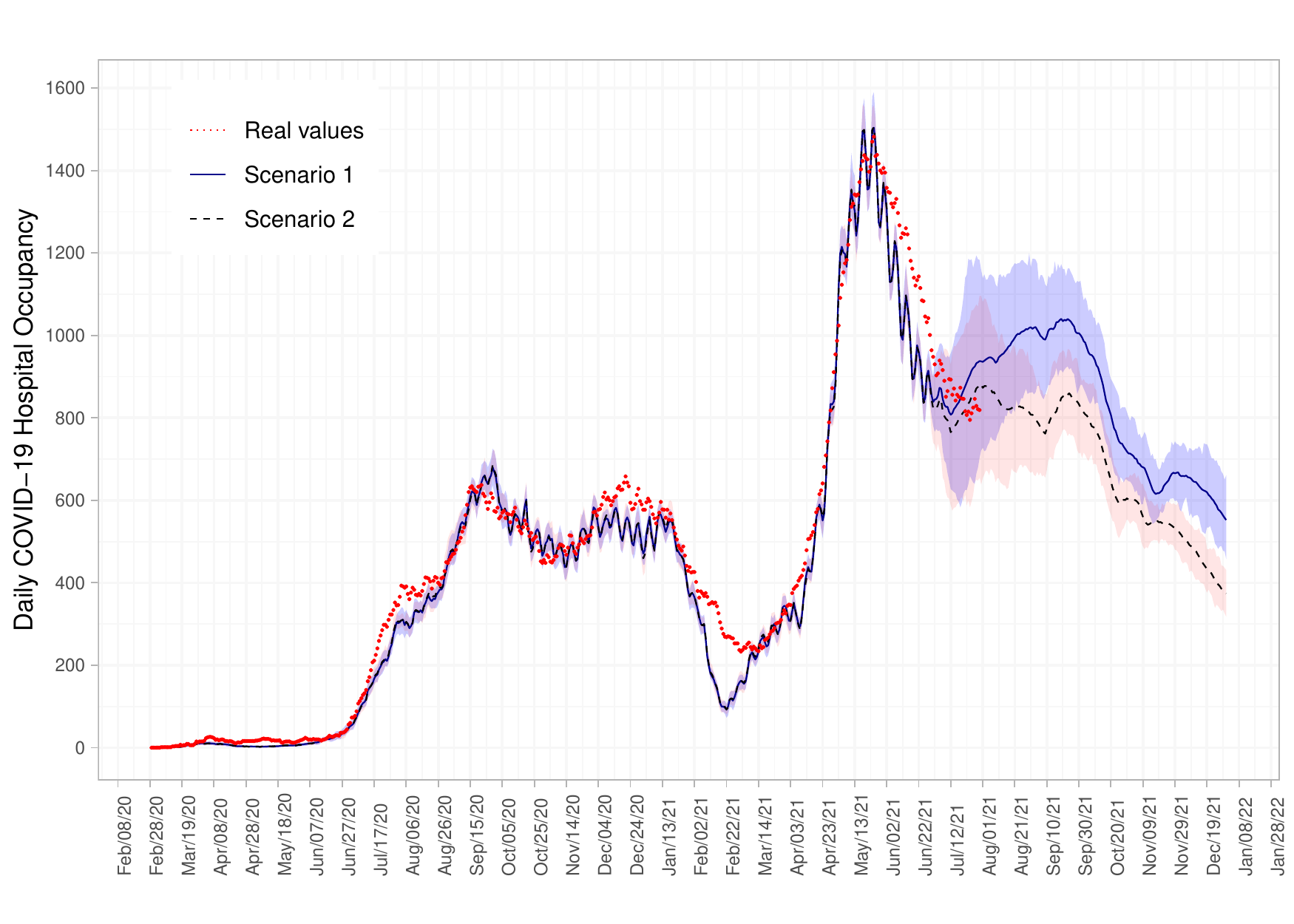}\\
\includegraphics[scale=0.4]{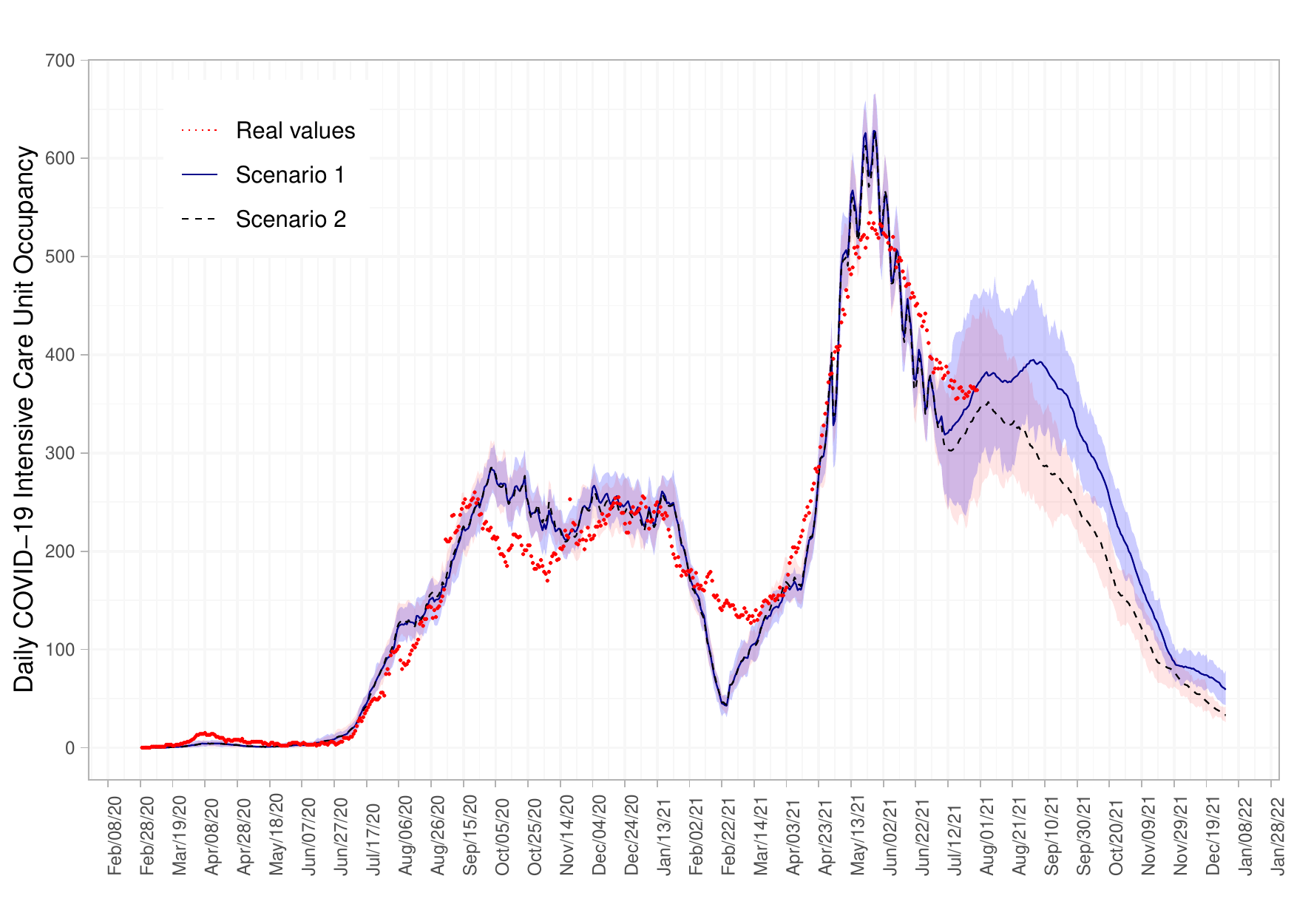}
\includegraphics[scale=0.4]{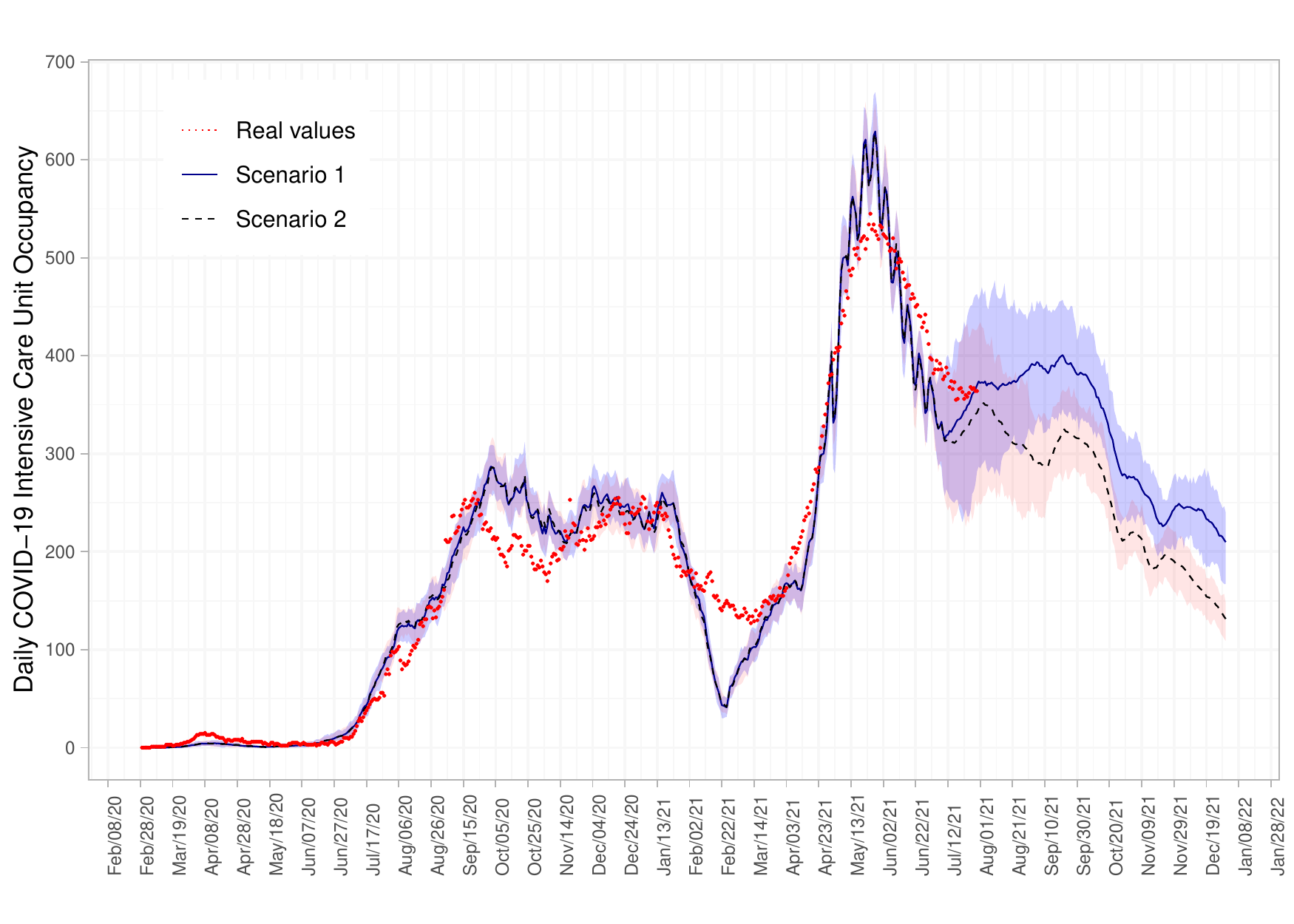}
\caption{{\bf Scenarios with Delta Variant}. Left panels correspond to the scenario without the introduction of Delta, and the right panels correspond to the results when the Delta variant becomes dominant over time. From top to bottom, cumulative Covid-19 confirmed cases, daily Covid-19 hospitalizations, and ICU Covid-19 occupancy.}
\label{fig:Esc2}
\end{figure}

We present the average of 50 simulations with a $95\%$ confidence interval (CI) implemented in Matlab R2020a~\cite{Matlab}. We used two remote servers:  (i) a Dell PowerEdge R740 with 64GB of RAM and two Intel®(R) Xeon®(R) Silver 4114 CPU @ 2.20GHz processors, and (ii) a Lenovo SR650, with two Intel® Xeon® Plata 4214, 2.20 GHz processors with 128GB of RAM. The outcome of our simulation is illustrated in Figure~\ref{fig:Esc2}. 

We discuss our results by taking two approaches. First, we compare variation in cases, deaths, and hospitalizations when the vaccination assumptions changed and the variants in circulation are the same by the rest of the year (scenario 1 and 2 in Table~\ref{tab:scenarios}, with Delta and current dominant variants). Then, we compare the impact of the introduction of the Delta variant with different vaccine coverage compared to the trends without this new SARS-CoV-2 variant. We also present the average base costs of the hospital stay in each scenario. All results are summarised in Table~\ref{tab:Results}. 

Among many others, low vaccine coverage translates into an economic impact for the hospital care system. According to data from the CCSS, in the country, Covid-19 patients are hospitalized from one to fifteen days in the medical ward and five to ten days in the ICU. This length of stay varies by age group. Considering the base cost of hospitalizations in CCSS, if a patients are hospitalized for an average of ten days in the Covid-19 ward, each patient would cost the Costa Rican public healthcare system 11,370 USD. Likewise, if patients are admitted to intensive care units for eight days in average, a patient would cost 11,600 USD, both without considering exams or procedures. We present the aggregated cost for basic hospital care taking these values into account. 

\subsection*{Scenario 1 and 2 considering the current variants as dominant for the rest of the year.}

A daily reduction in the administration of vaccines from 30,000 to 13,000 doses could lead to an increase in the cumulative confirmed cases by $6.20\%$ and total deaths by $7.9\% $ at the end of the year. Hospitalizations and ICU admissions from August 2 to December 31, 2021, could increase by $25.87\% $ and $28.30\%$ respectively ( from an average of 8,982 to an average of 11,306 for patients from the Covid-19 room and from 2,947 to 3,781 for ICU admissions). Low vaccine coverage could increase hospital stay costs by 26,423,880 USD in the ward and  9,674,400 USD in ICU.

\subsection*{Scenario 1 and 2 considering that Delta variant gradually becomes dominant over time.}

When we consider the circulation of the Delta variant and a reduction in daily vaccine administration, cumulative confirmed cases and cumulative deaths could increase by $6.75\%$ and $8.06\%$, respectively, for December 31, 2021. From August 2, 2021, to December 31, 2021, the new hospitalizations and ICUs admissions could rise by $22.66\%$ and $26.08\%$ (from 3,961 to 4,994 hospitalizations and from 7,762 to 7,183 ICU admissions), which in terms of hospital stay expenses correspond to an average increase of 31,654,080 USD in the ward and 11,982,800 USD in ICU.

\subsection*{Impact of the introduction of the Delta variant with different vaccine administration.}

We compared the trends with low acceptance and administration of the vaccine doses with and without the circulation of the Delta variant. We projected that the introduction of the new variant may increase the cumulative confirmed cases in $7.43\%$ and the total deaths in $9.04\%$ by the end of the year (from 572,955 to 615,574 cases and 7,095 to 7,762 deaths). Hospitalizations and ICU admissions from August 2, 2021, to December 31, 2021, are projected to increase by $33.30\%$ and $32.08\%$, respectively, from 11,306 to 15,071 hospitalizations in the ward which means a cumulative increase of 42,808,050 USD on base medical ward cost and from 3,781 to 4,994 ICU admissions, leading a cumulative increase on base ICU cost of 14,070,800 USD.

The same analysis in a high acceptance and vaccine administration leads to an increment of $6.89\%$ in cumulative confirmed cases, $9.3\%$ in cumulative deaths, $36.79\%$ in new hospitalizations (from 8,982 to 12,287), and $34.41\%$ in ICU admissions (from 2,947 to 3,961) representing an average increase in room and ICU stays of 37,577,850 USD and 11,762,400 USD, respectively.

The four scenarios showed an increase in hospital care demand by the end of June. This increment is regardless the introduction of Delta variant, as parameters that simulate its introduction are changed until the end of August. The increase in hospitalizations and ICU admissions reach a plateau that prolongs and decreases slowly with the progressive dominance of the Delta variant.

Our model projected that while maintaining a 12 week-interval between doses, a vaccine rolling out of 30,000 doses per day and high vaccine acceptance by the Costa Rica population, around $84.5\%$ of the target population would be vaccinated with at least one dose by mid-September, and approximately $75\%$ would be fully vaccinated by the end of the year. On the other hand, for a daily average vaccination of 13,000 doses per day with low acceptance of the vaccine, approximately $73.33\%$ of the target population would be vaccinated by mid-September, and $66.71\%$ would be fully vaccinated by the end of the year.

\bigskip
\begin{table}[!ht]
\centering
\begin{adjustbox}{width=1\textwidth}
\begin{tabular}{|>{\raggedright\arraybackslash}m{5.3cm}|
>{\centering\arraybackslash}m{3cm}|
>{\centering\arraybackslash}m{3cm}|
>{\centering\arraybackslash}m{3cm}|
>{\centering\arraybackslash}m{4cm}|
>{\centering\arraybackslash}m{3cm}|
>{\centering\arraybackslash}m{3cm}|
>{\centering\arraybackslash}m{3cm}|
>{\centering\arraybackslash}m{3cm}|}

\hline
\rowcolor{Gray1}
\hline
&\multicolumn{3}{|c|}{Current variants}&
\multicolumn{3}{|c|}{Delta variant} &
\multicolumn{2}{|c|}{Increase with the introduction of Delta }\\
\hline
\rowcolor{Gray}
\textbf{Description}&
\begin{minipage}[t]{\linewidth}%
\centering
\textbf{Vaccine strategy 1} \\
\textbf{13,000 doses per day} \\
\textbf{(95\% CI)}
\end{minipage}
&
\begin{minipage}[t]{\linewidth}%
\centering
\textbf{Vaccine strategy 2}\\
\textbf{30,000 doses per day}\\
\textbf{(95\% CI)}
\end{minipage}
&
\textbf{Increase  when vaccination rate is reduced} 
&
\begin{minipage}[t]{\linewidth}%
\centering
\textbf{Vaccine strategy 1}\\
\textbf{13,000 doses per day} \\
\textbf{(95\% CI)}
\end{minipage}
&
\begin{minipage}[t]{\linewidth}%
\centering
\textbf{Vaccine strategy 2} \\
\textbf{30,000 doses per day}\\
\textbf{(95\% CI)}
\end{minipage}
&
\textbf{Increase  when vaccination rate is reduced}
& 	
\textbf{Vaccine Strategy 1}	
&
\textbf{Vaccine Strategy 2}\\
\hline
Total positive cases by Dic 31, 2021 &	
\begin{minipage}[t]{\linewidth}%
\centering
572,955\\
(541337, 601308)
\end{minipage}
&
\begin{minipage}[t]{\linewidth}%
\centering
539,484\\
(511916, 569817) 
\end{minipage}
&6.20\% 
&
\begin{minipage}[t]{\linewidth}%
\centering
615,574\\
(589813, 640475) 
\end{minipage}
&
\begin{minipage}[t]{\linewidth}%
\centering
576,657\\
(550154, 599234) 
\end{minipage}
&	6.75\%	& 7.43\%	&6.89\%\\
\hline
New hospitalization from Aug 2 to Dic 31, 2021 &
\begin{minipage}[t]{\linewidth}%
\centering
11,306\\
(11318, 14189)
\end{minipage}
&
\begin{minipage}[t]{\linewidth}%
\centering
8,982 \\
(6815, 11393)
\end{minipage}
&	25.87\%
&
\begin{minipage}[t]{\linewidth}%
\centering
15,071\\
(11756, 18532)
\end{minipage}
&
\begin{minipage}[t]{\linewidth}%
\centering
12,287 \\
(9586, 15092)
\end{minipage}
&	22.66\%&	33.30\%&	36.79\%\\
\hline
New ICU admissions from Aug 2 to Dic 31, 2021 & 
\begin{minipage}[t]{\linewidth}%
\centering
3,781\\
(2542, 5092)
\end{minipage}
&
\begin{minipage}[t]{\linewidth}%
\centering
2,947\\
(1925, 4139)
\end{minipage}
&	28.30\%
&
\begin{minipage}[t]{\linewidth}%
\centering
4,994\\
(3 495, 6 593)
\end{minipage}
&
\begin{minipage}[t]{\linewidth}%
\centering
3,961 \\
(2691, 5336)
\end{minipage}
&	26.08\%	&32.08\% &	34.41\%\\
\hline
Total Deaths by Dic 31, 2021
&
\begin{minipage}[t]{\linewidth}%
\centering
7,095\\
(6 604,7 566)
\end{minipage}
&
\begin{minipage}[t]{\linewidth}%
\centering
6,572\\
(6273, 6936)
\end{minipage}
&	7.9\%
&
\begin{minipage}[t]{\linewidth}%
\centering
7,762\\
(7299, 8162)
\end{minipage}
&
\begin{minipage}[t]{\linewidth}%
\centering
7,183\\
(6847, 7567) 
\end{minipage}
&8.06\%	&9.4\%	&9.3\%\\
\hline
\begin{minipage}[t]{\linewidth}%
\centering
Average hospital charge  in ward\\
(10 days average stay)
\end{minipage}&
128,549,220 USD & 102,125,340 USD &  26,423,880 USD &
171,357,270 USD & 139,703,190 USD &  31,654,080 USD	&
42,808,050 USD &  37,577,850 USD\\

\hline
\begin{minipage}[t]{\linewidth}%
\centering
Average hospital charge in ICU\\
(8-days average stay)
\end{minipage}
&
43,859,600 USD & 34,185,200 USD&  9,674,400 USD&	
57,930,400 USD& 45,947,600 USD&  11,982,800 USD&
14,070,800 USD& 11,762,400 USD\\

\hline
Percentage of target people with at least one dose	& 
\begin{minipage}[t]{\linewidth}%
\centering
73.33\%\\
(73.21, 73.45)
\end{minipage}
&
\begin{minipage}[t]{\linewidth}%
\centering
84.50\%\\
(84.35, 84.65) 
\end{minipage}&	&	
\begin{minipage}[t]{\linewidth}%
\centering
72.89\% \\
(72.74, 73.02)	
\end{minipage}
&
\begin{minipage}[t]{\linewidth}%
\centering
84.50\%  \\
(84.37, 84.66) 
\end{minipage}
&	&	&	\\
\hline
Percentage of target people fully vaccinated &
\begin{minipage}[t]{\linewidth}%
\centering
66.71\%\\
(65.92, 67.53)
\end{minipage}
&
\begin{minipage}[t]{\linewidth}%
\centering
76.78\%\\
(75.69, 77.80)
\end{minipage}
&	 &	
\begin{minipage}[t]{\linewidth}%
\centering
65.67\%\\
(64.8, 66.57) 
\end{minipage}
& 
\begin{minipage}[t]{\linewidth}%
\centering
75.04\%\\
(74.02, 76.26)
\end{minipage}&	&	&	\\
\hline
\end{tabular}
\end{adjustbox}
\caption{{\bf Summary of results}. Hospital expenses are calculated based on a stay of 10 days in the medical ward and eight days in the ICU with an average cost per day of 707,540 colones (approximately 1,137 USD) in the medical ward and 902,641 colones (approximately 1,450 USD) in the ICU.}
\label{tab:Results}
\end{table}

\section*{Discussion}

With the emergence of SARS-CoV-2 variants influencing the transmissibility of the virus~\cite{van2021covid,walensky2021sars}, severity of the disease in younger population groups and the performance of available vaccines~\cite{bernal2021delta}, assessing probable future transmission scenarios and its resulting impact on health-care systems is at the forefront of health authorities and decision-makers around the world. By implementing a multilayer network model that simulates the introduction of the Delta variant in Costa Rica under two different scenarios of vaccine administration speed and population hesitance, we pursue to advocate in these efforts and generate evidence that may prove useful to inform interventions and preparations for public health authorities. 

By taking into account vaccination rates from December, 2020 to July, 2021 in Costa Rica, as well as levels of vaccine acceptance, our results suggest that reducing the daily application of first doses by an average of $57\%$ during a 5 month time-frame, Covid-19 related hospitalizations would experience an increase of $24\%$ on average, both in the scenarios of pre-Delta and Delta circulation, which translates in a $7.5\%$ increase mortality by the end of the study period. These results go in hand with previous studies that evidence the reduction in hospitalization and mortality with high efficacy mass vaccination campaigns~\cite{giordano2021modeling,borchering2021modeling,sah2021accelerated,alagoz2021impact}. 

Within a global scenario of huge disparities in access to Covid-19 vaccines~\cite{Vaccinei55:online}, middle-income countries, such as Costa Rica, are starting to see important increases in the supply of vaccines, now that worldwide production has strengthen and the richest nations have secured their supplies. However, we are now facing some of the same threats of vaccine hesitation among younger and healthier population groups. Furthermore, several countries have started to provide booster shots, which may slow down vaccine supplies worldwide one again. Our results provide evidence of the impact that implementing successful vaccination campaigns, including neutralizing disinformation campaigns, may have in terms of morbidity and mortality at the population level. We also provide initial evidence of the potential cost-effectiveness of such interventions, although further economic research is needed. 

In regard to the introduction of the Delta variant, we predicted that its entrance and gradual dominance in Costa Rica may increase hospitalizations on average $35\%$ and deaths on an average of $9.35\%$ in both the high and low vaccination scenarios. Although these figures are significant, it may contrast with some of the most abrupt increases seen in other jurisdictions, such as Israel or the UK, which may have several explanations. First, we assumed an introduction of the Delta on July 2021, thus, in our scenarios, Delta would not become relevant before August. In all four scenarios, Covid-19 hospitalizations and ICU admissions were projected to increase by the end of June 2021, regardless of the presence of Delta. This assumption is relevant because the magnitude of the projected outbreak depends not only in changes of transmissible and immune evasion of the virus, but also on the date of introduction of the variant, changes in social behavior, and vaccination coverage achieved at the time Delta becomes the dominants strain. Therefore, in the context of "delayed" Delta introduction and high vaccination coverage, even in the low vaccination rate scenario, our assumptions may have led to an optimistic result. 

All scenarios maintained current levels of non-pharmaceutical measures, including personal protection measures, mobility, and social gathering restrictions, which were only relaxed when $60\%$ of the population had a complete vaccination scheme, assuming changes in social behaviour and policy due to high levels of confidence. In a context where increasing sings of pandemic fatigue among the population~\cite{world2020pandemic} are arising, maintain these levels of compliance represents a challenge for public health authorities in the path for health, social and, economic recovery. Several of the sharp increases in cases seen in countries with high vaccination rates following the introduction of the Delta variant are in the context of non-pharmaceutical measures being almost completely lifted. 

Costa Rica is characterized for its near to universal health care coverage and relatively high historic investment in health promotion and prevention campaigns~\cite{saludpdf2:online,LibroPer43:online}. These fundamental factors have provided a relative protection against Covid-19 throughout the pandemic, flattening epidemiological curves and mitigating the effects of the pandemic in the country. 

The impact of a new variant can change geographically. Some previous variants of interest have made dramatic appearances in early locations and diminishing its impact in other jurisdictions where they reach dominance at a slower pace competing for dominance with other variants. This was for example the case of the Gamma variant outside South America. 

Results from the model also allowed us to estimate the increase in hospital care costs in each scenario, taking into account the base cost for hospital and ICU beds as reported by CCSS \cite{ccstarifa}. A slower vaccine administration pace translated into a total increase of 43,636,880 USD in base hospital care for the public health care system in the presence of the Delta variant. As the cost of diagnostics, specialized care and other procedures are not considered, the increase in costs for the social security in Costa Rica is expected to be substantially higher. Moreover, although a minority, several of these patients are managed in private hospitals, where the cost of base care is substantially higher. Additionally, if we take into account Costa Rica's near universal health care system, health care cost for the CCSS is expected to be substantially lower compared to other countries with more fragmented, unequal and limited health care coverage. 
It is worth noting that financing for the universal public health care system in Costa Rica comes from three sources: employers, workers, and the State, being the monthly social security contributions from its residents the main source of income. According to data from CCSS, income from social contributions from January 2020 to October 2020 decreased by 274,497,894 USD~\cite{657pdf6:online}, due to loss of jobs and income from the population, making our results even more relevant to consider for the allocation of resources within an already constrained system. 

There are limitations to using these mathematical and statistical tools that mostly lie in the model assumptions. For example, in this study, vaccines are applied randomly to susceptible and recovered individuals, regardless of their geographic location. Moreover, it is likely that the uptake of the vaccine is regionally correlated~\cite{office2020coronavirus}, leading to foci of high susceptibility in the population, which could act as small-scale outbreak sites reducing the effect of the immunity of the population~\cite{keeling2011effects}. Furthermore, the absorption of the vaccine may vary over time and geographically as the perception of risk changes~\cite{bish2011factors}. On the other hand, this assumption ignores that pre-symptomatic or asymptomatic people can receive the vaccine without knowing that they are carriers of the virus. Another relevant assumptions is related to the period that is needed to reach immunity after vaccination. This value is included only implicitly in the model, through the reduction of the percentages of protection against hospitalization. Studies report that protection after first dose begin approximately 12 days after first dose ~\cite{polack2020safety} and that it takes an estimated 14 days to reach maximum level of protection after second dose ~\cite{keehner2021sars,10.1093/cid/ciab554}. This period could become relevant when a high percentage of people is immunized and slightly shift our curves to the right.

The use of network models to forecast the effects of different vaccine strategies could be a valuable tool for decision-makers. These models have a flexible framework that allows for the incorporation of specific scenarios to study Covid-19 transmission dynamics. However, computational cost increases with the level of detail that is incorporated. This particular framework can be used in any country or region, however, the model is limited to specific information that, in many cases, is not available or difficult to access.

\bibliography{Biblio}{}
\bibliographystyle{abbrv}

\section*{Acknowledgements (not compulsory)}

The authors would like to thank the Research Center in Pure and Applied Mathematics and the School of Mathematics at Universidad de Costa Rica for their support during the preparation of this manuscript. They also thank the Ministry of Health for providing data and valuable information for this study. Thanks to Dr. María Dolores Pérez-Rosales, Representative of the Pan American Health Organization/World Health Organization in Costa Rica for her support and encouragement to pursue this work. 

\section*{Author contributions statement}

Y.G. and P.V., designed the study, performed simulations, analyzed data, and wrote the initial draft. G.M. and F.S designed the study, interpretation and analysis of the results, and edited the manuscript. J.C. coded the model and edited the manuscript. L.B. and T.R. designed the study and edited the manuscript. All authors reviewed the manuscript. G.M. is a staff member of the Pan American Health Organization. The author alone is responsible for the views expressed in this publication, and they do not necessarily represent the decisions or policies of the Pan American Health Organization.

\section*{Additional information}

\textbf{Codes}: All the codes and files necessary to reproduce the results presented in this document are available in the GitHub repository, \url{https://github.com/EpiMEC/NetworkModel_VaccinationScenarios}\\

\noindent \textbf{Competing interests}:
The authors declare no competing interests.

\end{document}